\documentclass[preprint]{ptephy}

\preprintnumber{KYUSHU-HET-198}

\usepackage{amssymb}
\usepackage{amsthm}
\usepackage{amsmath}
\usepackage{booktabs}
\usepackage{bbm}
\usepackage{mathtools}
\DeclareMathOperator{\Real}{Re}
\DeclareMathOperator{\tr}{tr}

\newcommand{\Slash}[1]{{\ooalign{\hfil/\hfil\crcr$#1$}}}
\numberwithin{equation}{section}

\begin{document}

\title{Infrared renormalon in $SU(N)$ QCD(adj.) on $\mathbb{R}^3\times S^1$}

\author{%
\name{\fname{Masahiro} \surname{Ashie}}{1},
\name{\fname{Okuto} \surname{Morikawa}}{1},
\name{\fname{Hiroshi} \surname{Suzuki}}{1,\ast},
\name{\fname{Hiromasa} \surname{Takaura}}{1},
and \name{\fname{Kengo} \surname{Takeuchi}}{1}
}

\address{%
\affil{1}{Department of Physics, Kyushu University
744 Motooka, Nishi-ku, Fukuoka, 819-0395, Japan}
\email{hsuzuki@phys.kyushu-u.ac.jp}
}

\date{\today}

\begin{abstract}
We study the infrared renormalon in the gluon condensate in the $SU(N)$ gauge
theory with $n_W$-flavor adjoint Weyl fermions (QCD(adj.))
on~$\mathbb{R}^3\times S^1$ with the $\mathbb{Z}_N$ twisted boundary
conditions. We rely on the so-called large-$\beta_0$ approximation as a
conventional tool to analyze the renormalon, in which only Feynman diagrams
that dominate in the large-$n_W$ limit are considered while the coefficient of
the vacuum polarization is set by hand to the one-loop beta
function~$\beta_0=11/3-2n_W/3$. In the large~$N$ limit within the
large-$\beta_0$ approximation, the W-boson, which acquires the twisted
Kaluza--Klein momentum, produces the renormalon ambiguity corresponding to the
Borel singularity at~$u=2$. This provides an example that the system in the
compactified space~$\mathbb{R}^3\times S^1$ possesses the renormalon ambiguity
identical to that in the uncompactified space~$\mathbb{R}^4$. We also discuss
the subtle issue that the location of the Borel singularity can change
depending on the order of two necessary operations.
\end{abstract}

\subjectindex{B00, B06, B32}
\maketitle

\section{Introduction}
\label{sec:1}
In the context of the resurgence program of asymptotically free quantum field
theories (for a review, see Ref.~\cite{Dunne:2015eaa} and the references cited
therein), the interesting possibility has been suggested that the ambiguity in
perturbation theory caused by the
infrared (IR) renormalon~\cite{tHooft:1977xjm,Beneke:1998ui}---a class of
Feynman diagrams whose amplitude grows factorially as a function of the order
of perturbation theory---is cancelled by the instability associated with a
semi-classical object called a bion~\cite{Argyres:2012vv,Argyres:2012ka,%
Dunne:2012ae,Dunne:2012zk}. This is analogous to the cancellation mechanism
between the ambiguity in perturbation theory around the trivial vacuum caused
by the proliferation of the number of Feynman diagrams and the instability
associated with an instanton--anti-instanton
pair~\cite{Bogomolny:1980ur,ZinnJustin:1981dx}. This possibility is very
intriguing because no one clearly knows what kind of non-perturbative effect
cancels the IR renormalon ambiguity. For a fully semi-classical understanding
of the physics of asymptotically-free quantum field theories along the
resurgence program, it appears essential to introduce a certain high-energy
scale such as a compactification radius of spacetime (see, for instance,
Ref.~\cite{Unsal:2007vu}). Thus, to reinforce the above picture on the IR
renormalon, the understanding of the IR renormalon in a compactified space such
as~$\mathbb{R}^{D-1}\times S^1$ is a basic premise.

The above picture has been examined fairly well in the two-dimensional (2D)
supersymmetric $\mathbb{C}P^{N-1}$ model~\cite{DAdda:1978dle} defined
on~$\mathbb{R}\times S^1$ with the $\mathbb{Z}_N$ twisted boundary
conditions~\cite{Dunne:2012ae,Dunne:2012zk}. In particular,
in~Ref.~\cite{Fujimori:2018kqp}, one-loop quantum corrections around the bion
configuration~\cite{Eto:2004rz,Eto:2006mz,Eto:2006pg,Bruckmann:2007zh,%
Brendel:2009mp,Bruckmann:2018rra} are explicitly computed and the associated
ambiguities are obtained, where the integration of the one-loop effective
action over quasi-collective coordinates is carried
out~\cite{Fujimori:2016ljw} following the Lefschetz thimble
method~\cite{Witten:2010cx,Cristoforetti:2012su,Fujii:2013sra}. In a recent
paper~\cite{Ishikawa:2019tnw}, on the other hand, the IR renormalon ambiguity
in the gluon condensate was determined in the leading order of the large-$N$
approximation~\cite{Coleman:1985rnk}. A very explicit calculation shows that a
Borel singularity at~$u=2$ (see below for this notion), which corresponds to
the IR renormalon in~$\mathbb{R}^2$, disappears for~$\mathbb{R}\times S^1$.
Instead of this, in the system on~$\mathbb{R}\times S^1$, an unfamiliar
renormalon singularity at~$u=3/2$ emerges. This is an unexpected result because
the IR renormalon singularity to be cancelled by the semi-classical bion has
been considered as the $u=2$ one. The observation
in~Ref.~\cite{Ishikawa:2019tnw} thus raises a question in the above
semi-classical picture on the IR renormalon.

As indicated in~Ref.~\cite{Ishikawa:2019tnw} and further discussed
in~Ref.~\cite{Ishikawa:2019oga}, the disappearance of the $u=2$ singularity and
the emergence of the $u=3/2$ singularity can be understood as a``shift'' of the
renormalon singularity under the
compactification~$\mathbb{R}^D\to\mathbb{R}^{D-1}\times S^1$. Moreover, it can
be seen that this is a very general phenomenon; it generally occurs provided
that the \emph{integrand\/} of the momentum integral in the ``renormalon
diagram'' for~$\mathbb{R}^{D-1}\times S^1$ is identical to that
for~$\mathbb{R}^D$, and that the Kaluza--Klein (KK) loop momentum is not
associated with the twisted boundary conditions; see below. The 2D
supersymmetric $\mathbb{C}P^{N-1}$ model in the large-$N$ limit satisfies these
prerequisites. See also Refs.~\cite{Eguchi:1982nm,Gross:1982at,%
Sulejmanpasic:2016llc} for a related ``volume independence'' property.

With the above observations, it is natural to repeat a similar analysis in 4D
gauge theories, in which the low-energy dynamics has been vigorously studied
aiming at a fully semi-classical understanding~\cite{Kovtun:2007py,%
Unsal:2007vu,Unsal:2007jx,Shifman:2008ja,Unsal:2008ch,Shifman:2009tp,%
Anber:2011de,Unsal:2012zj,Poppitz:2012sw,Poppitz:2012nz,Basar:2013sza,%
Poppitz:2013zqa,Anber:2013doa,Cherman:2014ofa,Misumi:2014raa,Anber:2014lba,%
Dunne:2016nmc,Cherman:2016hcd,Sulejmanpasic:2016llc,Yamazaki:2017ulc,%
Aitken:2017ayq,Tanizaki:2017qhf}. This is the motivation of the present paper.
We will study the IR renormalon in the gluon condensate in the $SU(N)$ gauge
theory with $n_W$-flavor adjoint Weyl fermions (QCD(adj.))
on~$\mathbb{R}^3\times S^1$ with the $\mathbb{Z}_N$ twisted boundary
conditions. Unlike the 2D $\mathbb{C}P^{N-1}$ model considered
in~Ref.~\cite{Ishikawa:2019tnw}, this system is much difficult to analyze and
does not allow a systematic treatment to study the renormalon. So, in this
paper, we rely on the so-called large-$\beta_0$ approximation, a somewhat ad
hoc but widely adopted prescription in studies of the renormalon in 4D gauge
theories~\cite{tHooft:1977xjm,Beneke:1994qe,Broadhurst:1993ru,Ball:1995ni}. In
the large-$\beta_0$ approximation, only Feynman diagrams that dominate in the
large-$n_W$ limit are considered, while the coefficient of the vacuum
polarization is set by hand to the one-loop coefficient of the beta function
(of the 't~Hooft coupling, see below),
\begin{equation}
   \beta_0=\frac{11}{3}-\frac{2}{3}n_W.
\label{eq:(1.1)}
\end{equation}
Despite the fact that the large-$\beta_0$ approximation is not a systematic
approach, this method is considered to be qualitatively reliable in gauge
theories on~$\mathbb{R}^4$ because the renormalon ambiguity obtained in this
approximation often has the same order of magnitude as the expected
non-perturbative effects (appearing in the context of the operator product
expansion).

In this large-$\beta_0$ approximation, we compute the one-loop effective action
to the quadratic order in the gauge field in a closed form for general~$N$.
From this, we obtain the gauge field propagator and then compute the gluon
condensate. The resulting expression is still rather complicated for explicit
analyses. Therefore, we further take an $N\to\infty$ limit\footnote{This is not
the genuine large-$N$ limit because we are working within the large-$\beta_0$
approximation which extracts a portion of the full set of Feynman diagrams.}
while the 't~Hooft coupling~$\lambda=g^2N$, where $g$ denotes the conventional
gauge coupling, and the one-loop dynamical scale,
\begin{equation}
   \Lambda\equiv\mu e^{-8\pi^2/(\beta_0\lambda)},
\label{eq:(1.2)}
\end{equation}
are kept fixed (here $\mu$ is the renormalization scale). The $S^1$ radius~$R$
is also kept fixed in this limit,
\begin{equation}
   \text{$\Lambda R=\text{const.}$ as $N\to\infty$}.
\label{eq:(1.3)}
\end{equation}
Then, we can show that terms peculiar to the compactified
space~$\mathbb{R}^3\times S^1$ in the gauge field propagator are suppressed.
This feature allows simpler analyses.

In this paper we adopt the following definitions in studying a factorially
divergent series. For the perturbative series of a quantity~$f(\lambda)$ in
the form
\begin{equation}
   f(\lambda)\sim\lambda\sum_{k=0}^\infty f_k
   \left(\frac{\beta_0\lambda}{16\pi^2}\right)^k,
\label{eq:(1.4)}
\end{equation}
we define the Borel transform by
\begin{equation}
   B(u)\equiv\sum_{k=0}^\infty\frac{f_k}{k!}u^k.
\label{eq:(1.5)}
\end{equation}
Then the Borel sum is defined by\footnote{In the 2D supersymmetric
$\mathbb{C}P^{N-1}$ model, we adopt the convention where the Borel integral is
given by~\cite{Ishikawa:2019tnw}
\begin{equation}
   f(\lambda)\equiv
   4\pi\int_0^\infty du\,B(u)\,e^{-4\pi u/\lambda},
\label{eq:(1.6)}
\end{equation}
such that the $u=2$ renormalon singularity also corresponds to twice the bion
action in the two-dimensional spacetime. (We note that $\beta_0=1$ for the 2D
supersymmetric $\mathbb{C}P^{N-1}$ model.)}
\begin{equation}
   f(\lambda)\equiv
   \frac{16\pi^2}{\beta_0}
   \int_0^\infty du\,B(u)\,e^{-16\pi^2u/(\beta_0\lambda)}.
\label{eq:(1.7)}
\end{equation}
If the perturbative coefficient~$f_k$ in~Eq.~\eqref{eq:(1.4)} grows
factorially, $f_k\sim b^{-k}k!$ as~$k\to\infty$, the Borel transform~$B(u)$
in~Eq.~\eqref{eq:(1.5)} develops a singularity at~$u=b$. If this singularity is
on the positive real $u$-axis (i.e.\ $b>0$), the Borel integral
in~Eq.~\eqref{eq:(1.7)} becomes ill-defined and produces an ambiguity
proportional to~$\sim e^{-16\pi^2 b/(\beta_0\lambda)}\propto\Lambda^{2b}$. In this
convention, the IR renormalon in the large-$\beta_0$ approximation produces
Borel singularities at positive integers $u=1$, $2$, \dots, for~$\mathbb{R}^4$.
On the other hand, since the classical action of the bion is~$16\pi^2/\lambda$
(when the constituent monopole-instanton and anti-monopole-instanton are
infinitely separated), the instability associated with the bion configuration
would produce singularities at~$u=n\beta_0$ with integer~$n$. Although, as it
stands, this does not coincide with the renormalon singularity
for~$\mathbb{R}^4$, it is conjectured~\cite{Argyres:2012vv} that quantum
corrections shift the bion contribution to~$u=1$, $2$, \dots.

In our analysis of the the gluon condensate in QCD(adj.)
on~$\mathbb{R}^3\times S^1$ (with the large-$N$ limit within the
large-$\beta_0$ approximation, as explained above), we find that the gluon
condensate suffers from the IR renormalon corresponding to the Borel
singularity at~$u=2$. The position of the singularity is \emph{identical\/} to
that of the system on~$\mathbb{R}^4$. Thus, the present system exhibits a
completely different property from the systems studied
in~Refs.~\cite{Ishikawa:2019tnw,Ishikawa:2019oga}, where the Borel singularity
at~$u=2$ for~$\mathbb{R}^2$ is shifted to~$u=3/2$ for~$\mathbb{R}\times S^1$.
This difference from the case of~Refs.~\cite{Ishikawa:2019tnw,Ishikawa:2019oga}
is attributed to the W-boson in the present system, which acquires the twisted
KK momentum as a consequence of the twisted boundary condition; the twisted KK
momentum is essential to keep the position of the Borel singularity unchanged
as we shall see.

To investigate the Borel singularity in the system on~$\mathbb{R}^3\times S^1$,
in fact, a careful treatment is required concerning how to take the large-$N$
limit for the perturbative series and the Borel transform. We find that the
result is sensitive to the order of the two operations: taking the large-$N$
limit and the construction of the Borel transform. The order of these
operations is not commutable: exchanging the order can lead to a completely
opposite conclusion, i.e.\ the emergence of the $u=3/2$ renormalon and the
disappearance of the $u=2$ renormalon.\footnote{This was the conclusion in the
first version of the present paper. We now consider that there are some
problems with this conclusion as discussed in~Sect.~\ref{sec:3}.} Since we are
interested in the divergence of the perturbative series in the system in the
large-$N$ limit, we should first obtain the perturbative coefficients with the
large-$N$ limit and then construct the Borel transform. This procedure leads to
the above conclusion that the $u=2$ renormalon exists.

Reference~\cite{Anber:2014sda} is a preceding study on the IR renormalon in the
$SU(N)$ QCD(adj.) in~$\mathbb{R}^3\times S^1$ with the $\mathbb{Z}_N$ twisted
boundary conditions. There, for $N=2$ and~$N=3$, the authors observed that the
one-loop vacuum polarization of the photon---the gauge field associated with
the Cartan subalgebra---does not have the logarithmic factor~$\sim\ln p^2$.
Since the IR renormalon is usually attributed to the existence of this
logarithmic factor, the authors concluded that there are no IR renormalons in
the $SU(N)$ QCD(adj.) in the compactified space~$\mathbb{R}^3\times S^1$ (at
least for $N=2$ and~$N=3$). This is not directly inconsistent with our result
in the present paper, because we consider the large-$N$ limit
of~Eq.~\eqref{eq:(1.3)}. Since we observe in this paper that the one-loop
vacuum polarization acquires the logarithmic factor~$\sim\ln p^2$, the
contribution of the compactification should depend on~$N$. It must be
interesting to clarify how the contribution of the compactification depends
on~$N$ and how small-$N$ cases and the large-$N$ case are connected.

This paper is organized as follows. In~Sect.~\ref{sec:2} we compute the
one-loop effective action to the quadratic order of the gauge field. We first
compute the contribution of the adjoint Weyl fermions. The twisted boundary
conditions, or equivalently the presence of a constant gauge potential, give
rise to some complications. Then, invoking the large-$\beta_0$ approximation,
we obtain the one-loop effective action and the gauge field propagator.
In~Sect.~\ref{sec:3}, using the gauge field propagator, we compute the gluon
condensate. We then determine the perturbative coefficients for the gluon
condensate in the large-$N$ limit. The corresponding Borel transform is then
constructed. We present these calculations for the photon and the W-boson parts
separately. We will also illustrate how the change of ordering of the large-$N$
limit and the construction of the Borel transform completely changes the
conclusion. Section~\ref{sec:4} is devoted to our conclusions.
In~Appendix~\ref{sec:A} we summarize our convention on the $SU(N)$ generators,
which is required for the computation in~Sect.~\ref{sec:2}.
In~Appendix~\ref{sec:B} we give a proof of the bounds that are crucial for the
above large-$N$ limit.

\section{One-loop effective action in the large-$\beta_0$ approximation}
\label{sec:2}
\subsection{Action and boundary conditions}
We assume that the spacetime is $\mathbb{R}^3\times S^1$ and the radius
of~$S^1$ is~$R$. The coordinates of~$\mathbb{R}^3$ are~$(x_0,x_1,x_2)$, and
that of~$S^1$ is~$x_3$; thus, $0\leq x_3<2\pi R$. The Euclidean action of the
$SU(N)$ QCD(adj.) is given by
\begin{equation}
   S=-\frac{N}{2\lambda_0}\int d^4x\,
   \tr\left[\Tilde{F}_{\mu\nu}(x)\Tilde{F}_{\mu\nu}(x)\right]
   -2\int d^4x\,\tr\left\{\Tilde{\Bar{\psi}}(x)\gamma_\mu\left[
   \partial_\mu\Tilde{\psi}(x)+[\Tilde{A}_\mu(x),\Tilde{\psi}(x)]
   \right]\right\}.
\label{eq:(2.1)}
\end{equation}
(The fields with a tilde are subject to the twisted boundary conditions, as
explained shortly.) Here, we have used the matrix notation with which
$\Tilde{A}_\mu(x)=-i\Tilde{A}_\mu^a(x)T^a$,
$\Tilde{\psi}(x)=-i\Tilde{\psi}^a(x)T^a$,
and~$\Tilde{\Bar{\psi}}(x)=-i\Tilde{\Bar{\psi}}^a(x)T^a$, where $T^a$ are
Hermitian $SU(N)$ generators in the fundamental representation, assuming the
normalization~$\tr(T^aT^b)=(1/2)\delta^{ab}$. Our convention for the $SU(N)$
generators is summarized in~Appendix~\ref{sec:A}. The field strength is defined
by
\begin{equation}
   \Tilde{F}_{\mu\nu}(x)=\partial_\mu\Tilde{A}_\nu(x)-\partial_\nu\Tilde{A}_\mu(x)
   +[\Tilde{A}_\mu(x),\Tilde{A}_\nu(x)],
\label{eq:(2.2)}
\end{equation}
and $\lambda_0$ is the bare 't~Hooft coupling that is related to the bare
gauge coupling~$g_0$ by~$\lambda_0=g_0^2N$. $\Tilde{\psi}(x)$
and~$\Tilde{\Bar{\psi}}(x)$ are $n_W$-flavor Weyl fermions and the summation
over the flavor index is suppressed for simplicity.

We assume that along~$S^1$ the above fields with a tilde obey the following
$\mathbb{Z}_N$-invariant twisted boundary conditions:
\begin{align}
   \Tilde{\psi}(x_0,x_1,x_2,x_3+2\pi R)
   &=\Omega\Tilde{\psi}(x_0,x_1,x_2,x_3)\Omega^{-1},
\notag\\
   \Tilde{\Bar{\psi}}(x_0,x_1,x_2,x_3+2\pi R)
   &=\Omega\Tilde{\Bar{\psi}}(x_0,x_1,x_2,x_3)\Omega^{-1},
\notag\\
   \Tilde{A}_\mu(x_0,x_1,x_2,x_3+2\pi R)
   &=\Omega\Tilde{A}_\mu(x_0,x_1,x_2,x_3)\Omega^{-1},
\label{eq:(2.3)}
\end{align}
where the $SU(N)$ element~$\Omega$ is defined by
\begin{equation}
   \Omega=e^{i\frac{2\pi}{N}\phi\cdot H}.
\label{eq:(2.4)}
\end{equation}
$H$ denotes the $SU(N)$ Cartan generator and the vector~$\phi$ is given by
\begin{equation}
   \phi_m\equiv2\sum_{j=1}^{N-1}(\mu^j)_m,
\label{eq:(2.5)}
\end{equation}
from the fundamental weights~$\mu^j=\sum_{k=1}^j\nu^k$ (here $\nu^k$ are
weights). See~Appendix~\ref{sec:A}. One can confirm that $\Omega$ is a
diagonal matrix with the diagonal elements
\begin{equation}
   e^{i\pi\frac{N+1}{N}}e^{-i\frac{2\pi}{N}j},\qquad j=1,2,\dots, N.
\label{eq:(2.6)}
\end{equation}
Since these diagonal elements are equally placed on the unit circle, the trace
of~$\Omega$ is invariant under the multiplication of the $\mathbb{Z}_N$ center
element, i.e.
\begin{equation}
   \tr(e^{i\frac{2\pi}{N}}\Omega)=\tr\Omega.
\label{eq:(2.7)}
\end{equation}
This is the origin of the name of the boundary conditions
in~Eq.~\eqref{eq:(2.3)}.

Instead of the above field variables with the twisted boundary conditions, it
is often convenient to use field variables which are periodic along~$S^1$. This
can be accomplished by substituting
\begin{align}
   \Tilde{\psi}(x_0,x_1,x_2,x_3)
   &=\Omega^{x_3/(2\pi R)}\psi(x_0,x_1,x_2,x_3)\Omega^{-x_3/(2\pi R)},
\notag\\
   \Tilde{\Bar{\psi}}(x_0,x_1,x_2,x_3)
   &=\Omega^{x_3/(2\pi R)}\Bar{\psi}(x_0,x_1,x_2,x_3)\Omega^{-x_3/(2\pi R)},
\notag\\
   \Tilde{A}_\mu(x_0,x_1,x_2,x_3)
   &=\Omega^{x_3/(2\pi R)}A_\mu(x_0,x_1,x_2,x_3)\Omega^{-x_3/(2\pi R)},
\label{eq:(2.8)}
\end{align}
where the variables on the right-hand side (without a tilde) are periodic
along~$S^1$. Note that under this substitution, the derivative acting on the
original variables with a tilde is translated into the covariant derivative
with respect to a constant gauge potential on the periodic field variables:
\begin{equation}
   \partial_\mu\to D_\mu^{(0)}\equiv\partial_\mu+[A_\mu^{(0)},\,],\qquad
   A_\mu^{(0)}\equiv i\frac{1}{RN}\phi\cdot H\delta_{\mu3}.
\label{eq:(2.9)}
\end{equation}


\subsection{Action in terms of component fields}
In what follows, we first extensively use the field variables that are periodic
in~$S^1$, the field variables on the right-hand side of~Eq.~\eqref{eq:(2.8)}.
We decompose these field variables in the Cartan--Weyl basis as
\begin{align}
   \psi(x)&=-i\sum_{\ell=1}^{N-1}\psi^\ell(x)H_\ell
   -i\sum_{m\neq n}\psi^{mn}(x)E_{mn},
\notag\\
   \Bar{\psi}(x)&=-i\sum_{\ell=1}^{N-1}\Bar{\psi}^\ell(x)H_\ell
   -i\sum_{m\neq n}\Bar{\psi}^{mn}(x)E_{mn},
\notag\\
   A_\mu(x)&=-i\sum_{\ell=1}^{N-1}A_\mu^\ell(x)H_\ell
   -i\sum_{m\neq n}A_\mu^{mn}(x)E_{mn}.
\label{eq:(2.10)}
\end{align}
See Appendix~\ref{sec:A} for our convention on the $SU(N)$ generators.
Throughout this paper, the gauge field~$A_\mu^\ell(x)$ is referred to as the
``photon'' and $A_\mu^{mn}(x)$ as the ``W-boson.'' Then, by using the
relations in~Eqs.~\eqref{eq:(A3)}, \eqref{eq:(A5)}, and
\begin{equation}
   \phi\cdot(\nu^m-\nu^n)
   =2\sum_{j=1}^{N-1}(\mu^j)\cdot(\nu^m-\nu^n)
   =-(m-n),
\label{eq:(2.11)}
\end{equation}
which follows from~Eq.~\eqref{eq:(A4)}, the action in~Eq.~\eqref{eq:(2.1)} in
terms of the periodic component fields is given by
\begin{align}
   S&=\frac{N}{4\lambda_0}\int d^4x\,
   \biggl\{
   \left[
   \partial_\mu A_\nu^\ell-\partial_\nu A_\mu^\ell
   -iA_\mu^{mn}A_\nu^{nm}(\nu^m-\nu^n)_\ell\right]
\notag\\
   &\qquad\qquad\qquad\qquad\qquad\qquad\qquad\qquad{}
   \times\left[
   \partial_\mu A_\nu^\ell-\partial_\nu A_\mu^\ell
   -iA_\mu^{pq}A_\nu^{qp}(\nu^p-\nu^q)_\ell\right]
\notag\\
   &\qquad\qquad\qquad{}
   +\biggl[
   \left(\partial_\mu-i\delta_{\mu3}\frac{m-n}{RN}\right)A_\nu^{mn}
   -\left(\partial_\nu-i\delta_{\nu3}\frac{m-n}{RN}\right)A_\mu^{mn}
\notag\\
   &\qquad\qquad\qquad\qquad{}
   -i(A_\mu^\ell A_\nu^{mn}-A_\mu^{mn}A_\nu^\ell)(\nu^m-\nu^n)_\ell
   -i\frac{1}{\sqrt{2}}
   (A_\mu^{m\ell}A_\nu^{\ell n}-A_\mu^{\ell n}A_\nu^{m\ell})\biggr]
\notag\\
   &\qquad\qquad\qquad\qquad{}
   \times\biggl[
   \left(\partial_\mu-i\delta_{\mu3}\frac{n-m}{RN}\right)A_\nu^{nm}
   -\left(\partial_\nu-i\delta_{\nu3}\frac{n-m}{RN}\right)A_\mu^{nm}
\notag\\
   &\qquad\qquad\qquad\qquad\qquad{}
   -i(A_\mu^p A_\nu^{nm}-A_\mu^{nm}A_\nu^p)(\nu^n-\nu^m)_p
   -i\frac{1}{\sqrt{2}}
   (A_\mu^{np}A_\nu^{pm}-A_\mu^{pm}A_\nu^{np})\biggr]
   \biggr\}
\notag\\
   &\qquad{}
   +\int d^4x\,\biggl\{
   \Bar{\psi}^m\Slash{\partial}\psi^m
   +\Bar{\psi}^{mn}
   \left(\Slash{\partial}
   -i\gamma_3\frac{n-m}{RN}\right)\psi^{nm}
\notag\\
   &\qquad\qquad\qquad\qquad{}
   -i\Bar{\psi}^{mn}
   \left[
   -\Slash{A}^\ell(\nu^m-\nu^n)_\ell\delta^{mq}\delta^{np}
   +\frac{1}{\sqrt{2}}(\Slash{A}^{np}\delta^{mq}-\Slash{A}^{qm}\delta^{np})
   \right]\psi^{pq}
\notag\\
   &\qquad\qquad\qquad\qquad{}
   -i\left[
   \Bar{\psi}^{mn}\Slash{A}^{nm}\psi^\ell(\nu^m-\nu^n)_\ell
   +\Bar{\psi}^\ell\Slash{A}^{mn}\psi^{nm}(\nu^m-\nu^n)_\ell
   \right]
   \biggr\},
\label{eq:(2.12)}
\end{align}
where we have taken the shift in~Eq.~\eqref{eq:(2.9)} into account. To this
gauge-invariant action we add the gauge-fixing term,
\begin{align}
   S_{\text{gf}}
   &=-\frac{N\xi_0}{\lambda_0}
   \int d^4x\,
   \tr\left[D_\mu^{(0)}A_\mu(x)D_\nu^{(0)}A_\nu(x)\right]
\notag\\
   &=\frac{N\xi_0}{2\lambda_0}\int d^4x\,\left[
   \partial_\mu A_\mu^\ell\partial_\nu A_\nu^\ell
   +\left(\partial_\mu-i\delta_{\mu3}\frac{m-n}{RN}\right)A_\mu^{mn}
   \left(\partial_\nu-i\delta_{\nu3}\frac{n-m}{RN}\right)A_\nu^{nm}\right],
\label{eq:(2.13)}
\end{align}
where $\xi_0$ is the bare gauge-fixing parameter and $D_\mu^{(0)}$ is the
covariant derivative in~Eq.~\eqref{eq:(2.9)}. Then, from the quadratic part of
the action~$S+S_{\text{gf}}$, we have free propagators of the periodic fields,
\begin{align}
   &\left\langle A_\mu^m(x)A_\nu^n(y)\right\rangle_0
   =\frac{\lambda_0}{N}\delta^{mn}
   \int\frac{d^3p}{(2\pi)^3}\,\frac{1}{2\pi R}\sum_{p_3}
   e^{ip(x-y)}
   \frac{1}{(p^2)^2}\left[(\delta_{\mu\nu}p^2-p_\mu p_\nu)
   +\frac{1}{\xi_0}p_\mu p_\nu\right],
\notag\\
   &\left\langle A_\mu^{mn}(x)A_\nu^{pq}(y)\right\rangle_0
\notag\\
   &=\frac{\lambda_0}{N}\delta^{mq}\delta^{np}
   \int\frac{d^3p}{(2\pi)^3}\,\frac{1}{2\pi R}\sum_{p_3}
   e^{ip(x-y)}
   \frac{1}{(p_{mn}^2)^2}\left[(\delta_{\mu\nu}p_{mn}^2-p_{mn,\mu}p_{mn,\nu})
   +\frac{1}{\xi_0}p_{mn,\mu}p_{mn,\nu}\right],   
\label{eq:(2.14)}
\end{align}
and
\begin{align}
   \left\langle\psi^m(x)\Bar{\psi}^n(y)\right\rangle_0
   &=\delta^{mn}\int\frac{d^3p}{(2\pi)^3}\frac{1}{2\pi R}\sum_{p_3}\,
   e^{ip(x-y)}
   \frac{1}{i\Slash{p}},
\notag\\
   \left\langle\psi^{mn}(x)\Bar{\psi}^{pq}(y)\right\rangle_0
   &=\delta^{mq}\delta^{np}
   \int\frac{d^3p}{(2\pi)^3}\frac{1}{2\pi R}\sum_{p_3}\,
   e^{ip(x-y)}
   \frac{1}{i\Slash{p}_{mn}}.
\label{eq:(2.15)}
\end{align}
In these expressions, $p_3$ denotes the KK momentum along~$S^1$, and thus
\begin{equation}
   p_3=\frac{n}{R},\qquad n\in\mathbb{Z}.
\label{eq:(2.16)}
\end{equation}
Also, we have introduced the \emph{twisted momentum},
\begin{equation}
   p_{mn,\mu}\equiv p_\mu-\delta_{\mu3}\frac{m-n}{RN}.
\label{eq:(2.17)}
\end{equation}
Note that the field components corresponding to the Cartan subalgebra do not
refer to the twisted momentum.

\subsection{One-loop effective action}
We now compute the vacuum polarization arising from one-loop radiative
corrections of the adjoint Weyl fermions.\footnote{This calculation is
required to construct the large-$\beta_0$ approximation.} This amounts to the
computation of the one-loop effective action of the gauge field to the
quadratic order arising from the Gaussian integration of the adjoint fermions.

Let us start with computing the part of the one-loop effective action that
contains $A_\mu^\ell(x)A_\nu^r(y)$. From the interaction terms
in~Eq.~\eqref{eq:(2.12)} and the free propagator in~Eq.~\eqref{eq:(2.15)}, we
have
\begin{align}
   {\mit\Gamma}^{(1)}
   &=-\frac{1}{2}
   \int d^4x\,d^4y\,
   A_\mu^\ell(x)A_\nu^r(y)
   \sum_{\substack{m\neq n\\1\leq m,n\leq N}}(\nu^m-\nu^n)_\ell(\nu^m-\nu^n)_r
   \int\frac{d^3p}{(2\pi)^3}\frac{1}{2\pi R}\sum_{p_3}
   e^{-ip(x-y)}
\notag\\
   &\qquad\qquad{}
   \times
   \frac{n_W}{2}\int\frac{d^3k}{(2\pi)^3}\frac{1}{2\pi R}\sum_{k_3}\tr
   \biggl[\frac{1}{i\Slash{k}-i\gamma_3(n-m)/(RN)}\gamma_\mu
\notag\\
   &\qquad\qquad\qquad\qquad\qquad\qquad\qquad\qquad{}
   \times\frac{1}{i(\Slash{k}-\Slash{p})-i\gamma_3(n-m)/(RN)}\gamma_\nu
   \biggr]+\dotsb.
\label{eq:(2.18)}
\end{align}
To this, we apply the identity
\begin{equation}
   \sum_{j=-\infty}^\infty e^{ik_32\pi Rj}
   =\frac{1}{R}\sum_{j=-\infty}^\infty\delta(k_3-j/R),
\label{eq:(2.19)}
\end{equation}
or
\begin{equation}
   \frac{1}{2\pi R}\sum_{j=-\infty}^\infty F(j/R)
   =\sum_{j=-\infty}^\infty\int\frac{dk_3}{2\pi}\,e^{ik_32\pi Rj}F(k_3),
\label{eq:(2.20)}
\end{equation}
to make the sum over $k_3$ the integrals~$\int dk_3$. After this, we can shift
the momentum variable as~$k_3\to k_3+(n-m)/(RN)$. Then, the trace over the
Dirac indices yields
\begin{align}
   {\mit\Gamma}^{(1)}
   &=n_W
   \int d^4x\,d^4y\,
   A_\mu^\ell(x)A_\nu^r(y)
   \sum_{\substack{m\neq n\\1\leq m,n\leq N}}(\nu^m-\nu^n)_\ell(\nu^m-\nu^n)_r
   \int\frac{d^3p}{(2\pi)^3}\frac{1}{2\pi R}\sum_{p_3}
   e^{-ip(x-y)}
\notag\\
   &\qquad\qquad{}
   \times\sum_{j=-\infty}^\infty e^{i(n-m)2\pi j/N}
   \int\frac{d^4k}{(2\pi)^4}\,e^{ik_32\pi Rj}
\notag\\
   &\qquad\qquad\qquad{}
   \times
   \int_0^1dx\,\frac{1}{(k^2-2xkp+xp^2)^2}
   \left[2k_\mu k_\nu-k_\mu p_\nu-p_\mu k_\nu-k(k-p)\delta_{\mu\nu}\right]
   +\dotsb.
\label{eq:(2.21)}
\end{align}
The summation over $m$ and~$n$ in this expression can be carried out by using
Eq.~\eqref{eq:(A1)} as
\begin{align}
   &(\sigma_{j,N})_{\ell r}
   \equiv
   \frac{1}{N}
   \sum_{\substack{m\neq n\\1\leq m,n\leq N}}(\nu^m-\nu^n)_\ell(\nu^m-\nu^n)_r
   e^{i(n-m)2\pi j/N}
\notag\\
   &=\begin{cases}
   \delta_{\ell r},\qquad\text{for $j=0\bmod N$},\\
   -\frac{1}{N}\frac{1}{\sqrt{\ell(\ell+1)r(r+1)}}\Real\left[
   \left(\dfrac{e^{-i\ell2\pi j/N}-1}{e^{-i2\pi j/N}-1}
   -\ell e^{-i\ell2\pi j/N}\right)
   \left(\dfrac{e^{ir2\pi j/N}-1}{e^{i2\pi j/N}-1}
   -r e^{ir2\pi j/N}\right)\right],\\
   \qquad\qquad\qquad\qquad\qquad\qquad\qquad\qquad\qquad\qquad\qquad\qquad
   \text{for $j\neq0\bmod N$}.
   \end{cases}
\label{eq:(2.22)}
\end{align}

In Eq.~\eqref{eq:(2.21)}, the term with~$j=0$ is ultraviolet (UV) divergent
while the terms with~$j\neq0$ are Fourier transforms and UV finite. We apply
dimensional regularization to the former by setting
$4\to D\equiv4-2\varepsilon$. Then the result of the momentum integrations is
\begin{align}
   {\mit\Gamma}^{(1)}
   &=\frac{1}{2}\frac{N}{16\pi^2}\frac{2}{3}n_W
   \int d^4x\,d^4y\,
   A_\mu^\ell(x)A_\nu^r(y)
   \int\frac{d^3p}{(2\pi)^3}\frac{1}{2\pi R}\sum_{p_3}
   e^{-ip(x-y)}
\notag\\
   &\qquad{}
   \times\biggl\{
   \delta_{\ell r}(p^2\delta_{\mu\nu}-p_\mu p_\nu)
   \left[
   \frac{1}{\varepsilon}+\ln(4\pi e^{-\gamma_E})
   +\ln\left(\frac{e^{5/3}}{p^2}\right)
   \right]
\notag\\
   &\qquad\qquad{}
   +12\sum_{j\neq0}(\sigma_{j,N})_{\ell r}
   \int_0^1dx\,e^{ixp_32\pi Rj}x(1-x)
\notag\\
   &\qquad\qquad{}
   \times\left[
   (p^2\delta_{\mu\nu}-p_\mu p_\nu)K_0(z)
   -(p^2\delta_{\mu3}\delta_{\nu3}
   -p_\mu p_3\delta_{\nu3}-p_\nu p_3\delta_{\mu3}+p_3^2\delta_{\mu\nu})K_2(z)
   \right]\biggr\}
\notag\\
   &\qquad{}+\dotsb,
\label{eq:(2.23)}
\end{align}
where $K_\nu(z)$ denotes the modified Bessel function of the second
kind\footnote{In obtaining Eq.~\eqref{eq:(2.23)}, one may use the relations
\begin{equation}
   K_0'(z)=-K_1(z),\qquad
   K_0(z)-K_2(z)=-\frac{2}{z}K_1(z),
\label{eq:(2.24)}
\end{equation}
and the relation following from integration by parts such as
\begin{equation}
   \int_0^1dx\,e^{ixp_32\pi Rj}
   \left[2ip_3\frac{\sqrt{x(1-x)}}{\sqrt{p^2}2\pi R|j|}K_1(z)
   -\frac{1-2x}{2\pi Rj}K_0(z)\right]=0,
\label{eq:(2.25)}
\end{equation}
because of~$zK_1'(z)+K_1(z)=-zK_0(z)$.} and
\begin{equation}
   z\equiv\sqrt{x(1-x)p^2}2\pi R|j|.
\label{eq:(2.26)}
\end{equation}

We can repeat a similar calculation for the term of the effective action
containing the combination~$A_\mu^{mn}(x)A_\nu^{pq}(y)$. After some calculation,
using $(\nu^m-\nu^n)^2=1$ for any fixed~$m\neq n$, we have
\begin{align}
   &{\mit\Gamma}^{(1)}
\notag\\
   &=\frac{1}{2}\frac{N}{16\pi^2}\frac{2}{3}n_W
   \int d^4x\,d^4y\,
   A_\mu^\ell(x)A_\nu^r(y)
   \int\frac{d^3p}{(2\pi)^3}\frac{1}{2\pi R}\sum_{p_3}
   e^{-ip(x-y)}
\notag\\
   &\qquad{}
   \times\biggl\{
   \delta_{\ell r}(p^2\delta_{\mu\nu}-p_\mu p_\nu)
   \left[
   \frac{1}{\varepsilon}+\ln(4\pi e^{-\gamma_E})
   +\ln\left(\frac{e^{5/3}}{p^2}\right)
   \right]
\notag\\
   &\qquad\qquad{}
   +12\sum_{j\neq0}(\sigma_{j,N})_{\ell r}
   \int_0^1dx\,e^{ixp_32\pi Rj}x(1-x)
\notag\\
   &\qquad\qquad{}
   \times\left[
   (p^2\delta_{\mu\nu}-p_\mu p_\nu)K_0(z)
   -(p^2\delta_{\mu3}\delta_{\nu3}
   -p_\mu p_3\delta_{\nu3}-p_\nu p_3\delta_{\mu3}+p_3^2\delta_{\mu\nu})K_2(z)
   \right]\biggr\}
\notag\\
   &\qquad{}
   +\frac{1}{2}\frac{N}{16\pi^2}\frac{2}{3}n_W
   \int d^4x\,d^4y\,
   A_\mu^{mn}(x)A_\nu^{nm}(y)
   \int\frac{d^3p}{(2\pi)^3}\frac{1}{2\pi R}\sum_{p_3}
   e^{-ip(x-y)}
\notag\\
   &\qquad{}
   \times\biggl\{
   (p^2\delta_{\mu\nu}-p_\mu p_\nu)
   \left[
   \frac{1}{\varepsilon}+\ln(4\pi e^{-\gamma_E})
   +\ln\left(\frac{e^{5/3}}{p^2}\right)
   \right]
\notag\\
   &\qquad\qquad{}
   +12\sum_{j\neq0,j=0\bmod N}
   \int_0^1dx\,e^{ixp_32\pi Rj}x(1-x)
\notag\\
   &\qquad\qquad{}
   \times\left[
   (p^2\delta_{\mu\nu}-p_\mu p_\nu)K_0(z)
   -(p^2\delta_{\mu3}\delta_{\nu3}
   -p_\mu p_3\delta_{\nu3}-p_\nu p_3\delta_{\mu3}+p_3^2\delta_{\mu\nu})K_2(z)
   \right]\biggr\}_{p\to p_{mn}}
\notag\\
   &\qquad{}+O(A^3).
\label{eq:(2.27)}
\end{align}
The explicit form of this expression depends on the assignment of the loop
momentum because the original integral in the $j=0$ term is UV divergent. The
different form corresponds to different regularization, and the difference can
be removed by a local counterterm. In~Eq.~\eqref{eq:(2.27)}, we adopted a
particular loop momentum assignment which leads to the simplest form. Note
that inside the last parentheses in~Eq.~\eqref{eq:(2.27)}, the momentum~$p$ is
replaced by~$p_{mn}$, the twisted momentum defined by~Eq.~\eqref{eq:(2.17)}. A
further calculation shows that no mixing term containing
$A_\mu^\ell(x)A_\nu^{mn}(y)$ arises. Equation~\eqref{eq:(2.27)} thus gives the
part of the effective action arising from one-loop radiative corrections of
$n_W$ Weyl fermions to the quadratic order in the gauge potential.

We now consider the large-flavor limit~$n_W\to\infty$ (with the
combination~$g^2n_W\propto\lambda n_W$ fixed), which is required as an
intermediate step in constructing the large-$\beta_0$ approximation. In this
approximation it is sufficient to consider only the fermion contribution to
the effective action of the gauge field as above, because radiative corrections
of the gauge field are subleading. In this way, we obtain the leading-order
result of the one-loop effective action in the large-$n_W$ limit
as~Eq.~\eqref{eq:(2.27)}, whose result is gauge invariant. Furthermore,
Eq.~\eqref{eq:(2.27)} is regarded as the same order as the classical action
[the sum of Eqs.~\eqref{eq:(2.12)} and~\eqref{eq:(2.13)}] for the gauge field
due to~$\lambda n_W=O(1)$. From~Eqs.~\eqref{eq:(2.12)}, \eqref{eq:(2.13)},
and~\eqref{eq:(2.27)}, we see that the effective action in this large-$n_W$
limit,
\begin{equation}
   S+S_{\text{gf}}+{\mit\Gamma}^{(1)},
\label{eq:(2.28)}
\end{equation}
is made finite by the following parameter renormalizations (in the
$\overline{\text{MS}}$ scheme):
\begin{equation}
   \lambda_0=\lambda
   \mu^{2\varepsilon}(4\pi e^{-\gamma_E})^{-\varepsilon}
   \mathcal{Z}^{-1},
   \qquad\xi_0=\xi\mathcal{Z}^{-1},\qquad
   \mathcal{Z}=1+\frac{\lambda}{16\pi^2}\left(-\frac{2}{3}n_W\right)
   \frac{1}{\varepsilon},
\label{eq:(2.29)}
\end{equation}
where $\mu$ is the renormalization scale and $\lambda$ denotes the
renormalized coupling at~$\mu$, i.e.\ $\lambda=\lambda(\mu)$.

In terms of these renormalized parameters, the effective action reads
\begin{align}
   &S+S_{\text{gf}}+{\mit\Gamma}^{(1)}
\notag\\
   &=\frac{N}{2\lambda}\int d^4x\,d^4y\,
   A_\mu^\ell(x)A_\nu^r(y)
   \int\frac{d^3p}{(2\pi)^3}\frac{1}{2\pi R}\sum_{p_3}
   e^{-ip(x-y)}
\notag\\
   &\qquad{}
   \times\biggl\{
   \delta_{\ell r}(p^2\delta_{\mu\nu}-p_\mu p_\nu)
   \left[1+\frac{\lambda}{16\pi^2}\frac{2}{3}n_W
   \ln\left(\frac{e^{5/3}\mu^2}{p^2}\right)
   \right]
   +\delta_{\ell r}\xi p_\mu p_\nu
\notag\\
   &\qquad\qquad{}
   +\frac{\lambda}{16\pi^2}\frac{2}{3}n_W12\sum_{j\neq0}(\sigma_{j,N})_{\ell r}
   \int_0^1dx\,e^{ixp_32\pi Rj}x(1-x)
\notag\\
   &\qquad\qquad{}
   \times\left[
   (p^2\delta_{\mu\nu}-p_\mu p_\nu)K_0(z)
   -(p^2\delta_{\mu3}\delta_{\nu3}
   -p_\mu p_3\delta_{\nu3}-p_\nu p_3\delta_{\mu3}+p_3^2\delta_{\mu\nu})K_2(z)
   \right]\biggr\}
\notag\\
   &\qquad{}
   +\frac{N}{2\lambda}
   \int d^4x\,d^4y\,
   A_\mu^{mn}(x)A_\nu^{nm}(y)
   \int\frac{d^3p}{(2\pi)^3}\frac{1}{2\pi R}\sum_{p_3}
   e^{-ip(x-y)}
\notag\\
   &\qquad{}
   \times\biggl\{
   (p^2\delta_{\mu\nu}-p_\mu p_\nu)
   \left[1+\frac{\lambda}{16\pi^2}\frac{2}{3}n_W
   \ln\left(\frac{e^{5/3}\mu^2}{p^2}\right)\right]
   +\xi p_\mu p_\nu
\notag\\
   &\qquad\qquad{}
   +\frac{\lambda}{16\pi^2}\frac{2}{3}n_W12\sum_{j\neq0,j=0\bmod N}
   \int_0^1dx\,e^{ixp_32\pi Rj}x(1-x)
\notag\\
   &\qquad\qquad{}
   \times\left[
   (p^2\delta_{\mu\nu}-p_\mu p_\nu)K_0(z)
   -(p^2\delta_{\mu3}\delta_{\nu3}
   -p_\mu p_3\delta_{\nu3}-p_\nu p_3\delta_{\mu3}+p_3^2\delta_{\mu\nu})K_2(z)
   \right]\biggr\}_{p\to p_{mn}}
\notag\\
   &\qquad{}+O(A^3).
\label{eq:(2.30)}
\end{align}

\subsection{Large-$\beta_0$ approximation}
Now, we consider the large-$\beta_0$ approximation, which is a somewhat ad hoc
way to include radiative corrections of the gauge field. In this approximation,
we first consider the large-$n_W$ limit as above. By this, we obtained the
gauge-invariant result of the one-loop effective action for the gauge field.
However, the large-$n_W$ limit breaks the asymptotic freedom and makes the
contribution of the gauge field to the vacuum polarization sub-dominant. To
remedy these points, the coefficient of the vacuum polarization is set by hand
to the one-loop coefficient of the beta function of the 't~Hooft coupling,
\begin{equation}
   -\frac{2}{3}n_W\to\beta_0=\frac{11}{3}-\frac{2}{3}n_W.
\label{eq:(2.31)}
\end{equation}
In this way, some part of the radiative corrections due to the gauge field is
supposed to be included.\footnote{It is worth noting that, in the
large-$\beta_0$ approximation, the leading logarithmic part of the
perturbative series is correctly obtained.}

Under Eq.~\eqref{eq:(2.31)}, the action given in~Eq.~\eqref{eq:(2.30)} is
changed to
\begin{align}
   &S+S_{\text{gf}}+{\mit\Gamma}^{(1)}
\notag\\
   &=\frac{N}{2\lambda}\int d^4x\,d^4y\,
   A_\mu^\ell(x)A_\nu^r(y)
   \int\frac{d^3p}{(2\pi)^3}\frac{1}{2\pi R}\sum_{p_3}
   e^{-ip(x-y)}
\notag\\
   &\qquad{}
   \times\biggl\{
   \delta_{\ell r}(p^2\delta_{\mu\nu}-p_\mu p_\nu)
   \left[1-\frac{\beta_0\lambda}{16\pi^2}
   \ln\left(\frac{e^{5/3}\mu^2}{p^2}\right)
   \right]
   +\delta_{\ell r}\xi p_\mu p_\nu
\notag\\
   &\qquad\qquad{}
   -\frac{\beta_0\lambda}{16\pi^2}12\sum_{j\neq0}(\sigma_{j,N})_{\ell r}
   \int_0^1dx\,e^{ixp_32\pi Rj}x(1-x)
\notag\\
   &\qquad\qquad{}
   \times\left[
   (p^2\delta_{\mu\nu}-p_\mu p_\nu)K_0(z)
   -(p^2\delta_{\mu3}\delta_{\nu3}
   -p_\mu p_3\delta_{\nu3}-p_\nu p_3\delta_{\mu3}+p_3^2\delta_{\mu\nu})K_2(z)
   \right]\biggr\}
\notag\\
   &\qquad{}
   +\frac{N}{2\lambda}
   \int d^4x\,d^4y\,
   A_\mu^{mn}(x)A_\nu^{nm}(y)
   \int\frac{d^3p}{(2\pi)^3}\frac{1}{2\pi R}\sum_{p_3}
   e^{-ip(x-y)}
\notag\\
   &\qquad{}
   \times\biggl\{
   (p^2\delta_{\mu\nu}-p_\mu p_\nu)
   \left[1-\frac{\beta_0\lambda}{16\pi^2}
   \ln\left(\frac{e^{5/3}\mu^2}{p^2}\right)\right]
   +\xi p_\mu p_\nu
\notag\\
   &\qquad\qquad{}
   -\frac{\beta_0\lambda}{16\pi^2}12\sum_{j\neq0,j=0\bmod N}
   \int_0^1dx\,e^{ixp_32\pi Rj}x(1-x)
\notag\\
   &\qquad\qquad{}
   \times\left[
   (p^2\delta_{\mu\nu}-p_\mu p_\nu)K_0(z)
   -(p^2\delta_{\mu3}\delta_{\nu3}
   -p_\mu p_3\delta_{\nu3}-p_\nu p_3\delta_{\mu3}+p_3^2\delta_{\mu\nu})K_2(z)
   \right]\biggr\}_{p\to p_{mn}}
\notag\\
   &\qquad{}+O(A^3).
\label{eq:(2.32)}
\end{align}
This is the effective action in the large-$\beta_0$ approximation.

\subsection{Gauge field propagator in the large-$\beta_0$ approximation}
We next obtain the gauge field propagator from the effective action
in~Eq.~\eqref{eq:(2.32)}. For this, it is convenient to introduce the
projection operators $\mathcal{P}_{\mu\nu}^T$ and~$\mathcal{P}_{\mu\nu}^L$
by~\cite{Anber:2014sda}
\begin{align}
   &\mathcal{P}_{ij}^T\equiv\delta_{ij}-\frac{p_ip_j}{p^2-p_3^2},\qquad
   \mathcal{P}_{i3}^T=\mathcal{P}_{3i}^T=\mathcal{P}_{33}^T\equiv0,
\notag\\
   &\mathcal{P}_{\mu\nu}^L\equiv\delta_{\mu\nu}-\frac{p_\mu p_\nu}{p^2}
   -\mathcal{P}_{\mu\nu}^T,
\label{eq:(2.33)}
\end{align}
where the Roman letters $i$, $j$, \dots, run only over $0$, $1$, and~$2$. These
satisfy $p_\mu\mathcal{P}_{\mu\nu}^T=\mathcal{P}_{\mu\nu}^Tp_\nu=
p_\mu\mathcal{P}_{\mu\nu}^L=\mathcal{P}_{\mu\nu}^Lp_\nu=0$ and, suppressing
Lorentz indices,
\begin{equation}
   \mathcal{P}^T\mathcal{P}^T=\mathcal{P}^T,\qquad
   \mathcal{P}^L\mathcal{P}^L=\mathcal{P}^L,\qquad
   \mathcal{P}^T\mathcal{P}^L=\mathcal{P}^L\mathcal{P}^T=0.
\label{eq:(2.34)}
\end{equation}
In terms of these projection operators, the effective action
in~Eq.~\eqref{eq:(2.32)} is expressed as
\begin{align}
   S+S_{\text{gf}}+{\mit\Gamma}^{(1)}
   &=\frac{N}{2\lambda}\int d^4x\,d^4y\,
   A_\mu^\ell(x)A_\nu^r(y)
   \int\frac{d^3p}{(2\pi)^3}\frac{1}{2\pi R}\sum_{p_3}
   e^{-ip(x-y)}
\notag\\
   &\qquad{}
   \times
   \left[p^2\mathcal{P}_{\mu\nu}^L(\delta^{\ell r}-L^{\ell r})
   +p^2\mathcal{P}_{\mu\nu}^T(\delta^{\ell r}-T^{\ell r})
   +\delta^{\ell r}\xi p_\mu p_\nu\right]
\notag\\
   &\qquad{}
   +\frac{N}{2\lambda}
   \int d^4x\,d^4y\,
   A_\mu^{mn}(x)A_\nu^{nm}(y)
   \int\frac{d^3p}{(2\pi)^3}\frac{1}{2\pi R}\sum_{p_3}
   e^{-ip(x-y)}
\notag\\
   &\qquad\qquad{}
   \times
   \left[p^2\mathcal{P}_{\mu\nu}^L(1-L)
   +p^2\mathcal{P}_{\mu\nu}^T(1-T)
   +\xi p_\mu p_\nu\right]_{p\to p_{mn}}
\notag\\
   &\qquad{}+O(A^3),
\label{eq:(2.35)}
\end{align}
with
\begin{align}
   L^{\ell r}&\equiv
   \frac{\beta_0\lambda}{16\pi^2}
   \Biggl\{
   \delta^{\ell r}\ln\left(\frac{e^{5/3}\mu^2}{p^2}\right)
\notag\\
   &\qquad\qquad\qquad{}
   +12\sum_{j\neq0}(\sigma_{j,N})_{\ell r}\int_0^1dx\,e^{ixp_32\pi Rj}x(1-x)
   \left[K_0(z)-K_2(z)\right]
   \Biggr\},
\notag\\
   T^{\ell r}&\equiv
   \frac{\beta_0\lambda}{16\pi^2}
   \Biggl\{
   \delta^{\ell r}
   \ln\left(\frac{e^{5/3}\mu^2}{p^2}\right)
\notag\\
   &\qquad\qquad\qquad{}
   +12\sum_{j\neq0}(\sigma_{j,N})_{\ell r}\int_0^1dx\,e^{ixp_32\pi Rj}x(1-x)
   \left[K_0(z)-\frac{p_3^2}{p^2}K_2(z)\right]
   \Biggr\},
\notag\\
   L&\equiv
   \frac{\beta_0\lambda}{16\pi^2}
   \Biggl\{\ln\left(\frac{e^{5/3}\mu^2}{p^2}\right)
\notag\\
   &\qquad\qquad\qquad{}
   +12\sum_{j\neq0,j=0\bmod N}\int_0^1dx\,e^{ixp_32\pi Rj}x(1-x)
   \left[K_0(z)-K_2(z)\right]
   \Biggr\},
\notag\\
   T&\equiv
   \frac{\beta_0\lambda}{16\pi^2}
   \Biggl\{\ln\left(\frac{e^{5/3}\mu^2}{p^2}\right)
\notag\\
   &\qquad\qquad\qquad{}
   +12\sum_{j\neq0,j=0\bmod N}\int_0^1dx\,e^{ixp_32\pi Rj}x(1-x)
   \left[K_0(z)-\frac{p_3^2}{p^2}K_2(z)\right]
   \Biggr\},
\label{eq:(2.36)}
\end{align}
where the variable~$z$ is defined by~Eq.~\eqref{eq:(2.26)}. From this
expression, we see that the propagators of the gauge field \emph{with the
twisted boundary conditions in~Eq.~\eqref{eq:(2.3)}\/} are given
by\footnote{For this, we have to recall the relation in~Eq.~\eqref{eq:(2.8)}.}
\begin{align}
   &\left\langle\Tilde{A}_\mu^\ell(x)\Tilde{A}_\nu^r(y)\right\rangle
\notag\\
   &=\frac{\lambda}{N}\int\frac{d^3p}{(2\pi)^3}
   \frac{1}{2\pi R}\sum_{p_3}
\notag\\
   &\qquad\qquad{}
   \times e^{ip(x-y)}
   \frac{1}{(p^2)^2}
   \left\{
   \left[(1-L)^{-1}\right]^{\ell r}p^2\mathcal{P}_{\mu\nu}^L
   +\left[(1-T)^{-1}\right]^{\ell r}p^2\mathcal{P}_{\mu\nu}^T
   +\delta^{\ell r}\frac{1}{\xi}p_\mu p_\nu
   \right\},
\notag\\
   &\left\langle\Tilde{A}_\mu^{mn}(x)\Tilde{A}_\nu^{pq}(y)\right\rangle
\notag\\
   &=\frac{\lambda}{N}\delta^{mq}\delta^{np}
   \int\frac{d^3p}{(2\pi)^3}
   \frac{1}{2\pi R}\sum_{p_3}
\notag\\
   &\qquad\qquad{}
   \times\left\{e^{ip(x-y)}
   \frac{1}{(p^2)^2}
   \left[
   (1-L)^{-1}p^2\mathcal{P}_{\mu\nu}^L
   +(1-T)^{-1}p^2\mathcal{P}_{\mu\nu}^T
   +\frac{1}{\xi}p_\mu p_\nu
   \right]\right\}_{p\to p_{mn}}.
\label{eq:(2.37)}
\end{align}
In the last expression, the twisted momentum~$p_{mn}$ is substituted, which is
defined by~Eq.~\eqref{eq:(2.17)}.

\section{Borel singularity in the gluon condensate in the large-$N$ limit}
\label{sec:3}
In this section we compute the gluon condensate in the large-$\beta_0$
approximation by using~Eq.~\eqref{eq:(2.37)}, and determine the perturbative
coefficients for the gluon condensate under the large-$N$ limit
of~Eq.~\eqref{eq:(1.3)}. We then construct the corresponding Borel transform.
The IR renormalon ambiguity associated with the gluon condensate
in~$\mathbb{R}^4$ has been studied in~Refs.~\cite{Bali:2014sja,Bali:2015cxa,%
Lee:2010hd,Lee:2015bci,DelDebbio:2018ftu,Horsley:2012ra}.

\begin{figure}
\centering
\includegraphics[width=0.2\textwidth,clip]{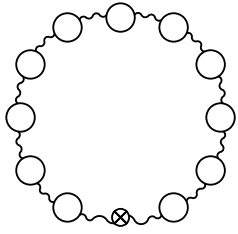}
\caption{The Feynman diagram dominating the gluon condensate~\eqref{eq:(3.1)}
in the large-$\beta_0$ approximation. The gauge field propagators
in~Eq.~\eqref{eq:(2.37)}, which are given by a chain of vacuum polarizations,
are used to contract two gauge fields
in~$\tr(\Tilde{F}_{\mu\nu}\Tilde{F}_{\mu\nu})$.}
\label{fig:1}
\end{figure}

In the large-$\beta_0$ approximation, the gluon condensate is computed as
(see~Fig.~\ref{fig:1})
\begin{align}
   &\left\langle\tr(\Tilde{F}_{\mu\nu}\Tilde{F}_{\mu\nu})\right\rangle
\notag\\
   &=-\frac{1}{2}
   \left\langle
   (\partial_\mu\Tilde{A}_\nu^\ell-\partial_\nu\Tilde{A}_\mu^\ell)^2
   \right\rangle
   -\frac{1}{2}
   \left\langle
   (\partial_\mu\Tilde{A}_\nu^{mn}-\partial_\nu\Tilde{A}_\mu^{mn})
   (\partial_\mu\Tilde{A}_\nu^{nm}-\partial_\nu\Tilde{A}_\mu^{nm})
   \right\rangle
\notag\\
   &=-\frac{\lambda}{N}
   \int\frac{d^3p}{(2\pi)^3}\,\frac{1}{2\pi R}\sum_{p_3}
   \sum_{\ell=1}^{N-1}\left\{
   \left[(1-L)^{-1}\right]^{\ell\ell}
   +2\left[(1-T)^{-1}\right]^{\ell\ell}
   \right\}
\notag\\
   &\qquad{}
   -\frac{\lambda}{N}
   \int\frac{d^3p}{(2\pi)^3}\,\frac{1}{2\pi R}\sum_{p_3}
   \sum_{\substack{m\neq n\\1\leq m,n\leq N}}
   \left[(1-L)^{-1}+2(1-T)^{-1}\right]_{p\to p_{mn}}.
\label{eq:(3.1)}
\end{align}
In the last expression, the first line corresponds to the contribution of the
photon, the gauge field associated with the Cartan subalgebra, whereas the
second line does to the W-boson which acquires the twisted momentum~$p_{mn}$
due to the twisted boundary conditions. We treat these contributions
separately.

\subsection{Contribution of the photon}
Since the functions~$L^{\ell r}$ and~$T^{\ell r}$ in~Eq.~\eqref{eq:(2.36)} are
$O(\lambda)$, we obtain the perturbative expansion of the gluon condensate for
the photon part as
\begin{equation}
   \left\langle\tr(\Tilde{F}_{\mu\nu}\Tilde{F}_{\mu\nu})
   \right\rangle_{\text{photon}}
   =-\frac{\lambda}{N}\sum_{k=0}^\infty
   \int\frac{d^3p}{(2\pi)^3}\,\frac{1}{2\pi R}\sum_{p_3}
   \sum_{\ell=1}^{N-1}\left[
   (L^k)^{\ell\ell}
   +2(T^k)^{\ell\ell}
   \right],
\label{eq:(3.2)}
\end{equation}
from which the $k$th perturbative coefficient can be read off.
In~Appendix~\ref{sec:B} we show that, in the large-$N$ limit
of~Eq.~\eqref{eq:(1.3)},
\begin{equation}
   \sum_{j\neq0}\sigma_{j,N}\int dx\,e^{ixp_32\pi Rj}x(1-x)K_\nu(z)=O(1/N),
\label{eq:(3.3)}
\end{equation}
for $\nu=0$ or~$2$. Therefore, the second terms in~$L^{\ell r}$ and
in~$T^{\ell r}$ of~Eq.~\eqref{eq:(2.36)} give only sub-dominant contributions in
the large-$N$ limit.\footnote{The bounds for the finite volume corrections get
larger at lower-energy region, as shown
in~Eqs.~\eqref{eq:(B7)}--\eqref{eq:(B10)}. Thus, the finite volume corrections
may cause IR divergences in perturbative coefficients. However, we regard such
divergence as the subleading effect in terms of large-$N$. The same
identification is applied to the calculation of the contribution from the
W-boson.} Thus, the $k$th perturbative coefficient $f_k$ [defined as
in~Eq.~\eqref{eq:(1.4)}] in the large-$N$ limit is given by
\begin{equation}
   (f_k)_{\text{photon}}
   =-3\int\frac{d^3p}{(2\pi)^3}\frac{1}{2\pi R}\sum_{p_3}
   \left[\ln\left(\frac{e^{5/3}\mu^2}{p^2}\right)\right]^k,
\label{eq:(3.4)}
\end{equation}
which is~$O(N^0)$.

We construct the corresponding Borel transform [cf.\ Eq.~\eqref{eq:(1.5)}] to
investigate the large-order behavior:
\begin{align}
   B(u)_{\text{photon}}
   &=-3\int\frac{d^3p}{(2\pi)^3}\frac{1}{2\pi R}\sum_{p_3}
   \left(\frac{e^{5/3}\mu^2}{p^2}\right)^u
\notag\\
   &=-3\sum_{j=-\infty}^\infty
   \int\frac{d^4p}{(2\pi)^4}\,e^{ip_32\pi Rj}
   \left(\frac{e^{5/3}\mu^2}{p^2}\right)^u,
\label{eq:(3.5)}
\end{align}
where we have used Eq.~\eqref{eq:(2.20)}. In this expression, the term
with~$j=0$ is UV divergent and we introduce a UV cutoff~$q>0$ to the momentum
integral, $|p|\leq q$. The $j\neq0$ terms are Fourier transforms and UV
convergent. Then, the momentum integration yields
\begin{equation}
   B(u)_{\text{photon}}
   =\frac{3}{16\pi^2}(e^{5/3}\mu^2)^u
   \left[
   (q^2)^{2-u}\frac{1}{u-2}
   -2(\pi^2R^2)^{u-2}\frac{{\mit\Gamma}(2-u)}{{\mit\Gamma}(u)}\zeta(4-2u)
   \right].
\label{eq:(3.6)}
\end{equation}
The only singularity of this function is given by the simple pole at~$u=3/2$:
\begin{equation}
   B(u)_{\text{photon}}
   \stackrel{u\sim3/2}{\sim}
   \frac{3}{16\pi^2}(e^{5/3}\mu^2)^{3/2}
   2(\pi^2R^2)^{-1/2}\frac{1}{u-3/2}.
\label{eq:(3.7)}
\end{equation}
We again note that this is~$O(N^0)$. In fact, the photon part satisfies the
prerequisites for the analysis of~Ref.~\cite{Ishikawa:2019oga} and the general
argument therein indicates the singularity at~$u=3/2$, as a consequence of the
shift of the singularity by~$-1/2$ in the Borel $u$-plane.

\subsection{Contribution of the W-boson}
The perturbative expansion of the W-boson part is given by
\begin{equation}
   \left\langle\tr(\Tilde{F}_{\mu\nu}\Tilde{F}_{\mu\nu})
   \right\rangle_{\text{W-boson}}
   =-\frac{\lambda}{N}\sum_{k=0}^\infty
   \int\frac{d^3p}{(2\pi)^3}\,\frac{1}{2\pi R}\sum_{p_3}
   \sum_{\substack{m\neq n\\1\leq m,n\leq N}}
   (L^k+2T^k)_{p\to p_{mn}}.
\label{eq:(3.8)}
\end{equation}
Since we can show that (Appendix~\ref{sec:B})
\begin{equation}
   \sum_{j\neq0,j=0\bmod N}\int dx\,e^{ixp_32\pi Rj}x(1-x)K_\nu(z)=O(1/N^3),
\label{eq:(3.9)}
\end{equation}
the second terms of~$L$ and~$T$ in~Eq.~\eqref{eq:(2.36)} again give only
sub-dominant contribution in the large-$N$ limit. Hence, in the large-$N$
limit, we obtain the $k$th perturbative coefficient $f_k$ [defined as
in~Eq.~\eqref{eq:(1.4)}] as
\begin{align}
   (f_k)_{\text{W-boson}}
   &=-3\frac{1}{N}
   \int\frac{d^3p}{(2\pi)^3}\frac{1}{2\pi R}\sum_{p_3}
   \sum_{\substack{m\neq n\\1\leq m,n\leq N}}
   \left[\ln\left(\frac{e^{5/3}\mu^2}{p_{mn}^2}\right)\right]^k
\notag\\
   &=-3\sum_{j=-\infty}^\infty
   \int\frac{d^4p}{(2\pi)^4}\,e^{ip_32\pi Rj}\frac{1}{N}
   \sum_{\substack{m\neq n\\1\leq m,n\leq N}}e^{i(m-n)2\pi j/N}
   \left[\ln\left(\frac{e^{5/3}\mu^2}{p^2}\right)\right]^k,
\label{eq:(3.10)}
\end{align}
where we have used Eqs.~\eqref{eq:(2.20)} and~\eqref{eq:(2.17)}, and shifted
the momentum~$p_3\to p_3+(m-n)/(RN)$. Now, we note that
\begin{equation}
   \frac{1}{N}
   \sum_{\substack{m\neq n\\1\leq m,n\leq N}}e^{i(m-n)2\pi j/N}
   =\begin{cases}
   N-1,&\text{for $j=0\mod N$},\\
   -1,&\text{for $j\neq 0\mod N$},\\
   \end{cases}
\label{eq:(3.11)}
\end{equation}
and thus
\begin{align}
   (f_k)_{\text{W-boson}}
   &=-3\sum_{j=-\infty}^\infty
   \int\frac{d^4p}{(2\pi)^4}\,e^{ip_32\pi RNj}
   \left[(N-1)-(-1)\right]
   \left[\ln\left(\frac{e^{5/3}\mu^2}{p^2}\right)\right]^k
\notag\\
   &\qquad{}
   -3\sum_{j=-\infty}^\infty
   \int\frac{d^4p}{(2\pi)^4}\,e^{ip_32\pi Rj}(-1)
   \left[\ln\left(\frac{e^{5/3}\mu^2}{p^2}\right)\right]^k.
\label{eq:(3.12)}
\end{align}
The second line on the right-hand side precisely cancels the contribution of
the photon in~Eq.~\eqref{eq:(3.4)}. For the first line, we apply
Eq.~\eqref{eq:(2.20)} in an opposite way
(cf.\ Ref.~\cite{Sulejmanpasic:2016llc}):
\begin{align}
   (f_k)_{\text{W-boson}}
   &=-3N
   \int\frac{d^3p}{(2\pi)^3}\,\frac{1}{2\pi RN}\sum_{p_3=n/(RN)}
   \left[\ln\left(\frac{e^{5/3}\mu^2}{p^2}\right)\right]^k
\notag\\
   &\qquad{}
   -3\sum_{j=-\infty}^\infty
   \int\frac{d^4p}{(2\pi)^4}\,e^{ip_32\pi Rj}(-1)
   \left[\ln\left(\frac{e^{5/3}\mu^2}{p^2}\right)\right]^k.
\label{eq:(3.13)}
\end{align}
Remarkably, in the first term of the right-hand side, the effective radius of
the compactified direction becomes~$RN$ as a consequence of the twisted
boundary conditions. Hence, $S^1$ is effectively decompactified in the
large-$N$ limit, and the first term is reduced to the expression in the
uncompactified~$\mathbb{R}^4$:
\begin{align}
   (f_k)_{\text{W-boson}}
   &=-3N
   \int\frac{d^4p}{(2\pi)^4}\,
   \left[\ln\left(\frac{e^{5/3}\mu^2}{p^2}\right)\right]^k
\notag\\
   &\qquad{}
   -3\sum_{j=-\infty}^\infty
   \int\frac{d^4p}{(2\pi)^4}\,e^{ip_32\pi Rj}(-1)
   \left[\ln\left(\frac{e^{5/3}\mu^2}{p^2}\right)\right]^k,
\label{eq:(3.14)}
\end{align}
where the sum over the KK momentum has been replaced by an integral in the
$N\to\infty$ limit~\cite{Sulejmanpasic:2016llc}.

From these perturbative coefficients, we obtain the Borel transform
[cf.\ Eq.~\eqref{eq:(1.5)}] as
\begin{align}
   B(u)_{\text{W-boson}}
   &=-3N\int\frac{d^4p}{(2\pi)^4}\left(\frac{e^{5/3}\mu^2}{p^2}\right)^u
   -3\sum_{j=-\infty}^\infty
   \int\frac{d^4p}{(2\pi)^4}\,e^{ip_32\pi Rj}(-1)
   \left(\frac{e^{5/3}\mu^2}{p^2}\right)^u
\notag\\
   &=\frac{3N}{16\pi^2}(e^{5/3}\mu^2)^u(q^2)^{2-u}\frac{1}{u-2}
   -B(u)_{\text{photon}}.
\label{eq:(3.15)}
\end{align}
This has the singularity at~$u=2$; this position coincides with that of the
uncompactified spacetime $\mathbb{R}^4$. We note that the contribution from
the W-boson is of~$O(N)$.

\subsection{IR renormalon in the gluon condensate}
As the sum of Eqs.~\eqref{eq:(3.6)} and~\eqref{eq:(3.15)}, we have
\begin{equation}
   B(u)=B(u)_{\text{photon}}+B(u)_{\text{W-boson}}
   =\frac{3N}{16\pi^2}(e^{5/3}\mu^2)^u(q^2)^{2-u}\frac{1}{u-2}.
\label{eq:(3.16)}
\end{equation}
Therefore, the gluon condensate in the present system suffers from the IR
renormalon ambiguity corresponding to~$u=2$. Through the Borel sum
in~Eq.~\eqref{eq:(1.7)}, this pole singularity produces the ambiguity,
\begin{equation}
   \frac{3N}{\beta_0}(e^{5/3}\mu^2)^2e^{-32\pi^2/(\beta_0\lambda)}(\mp\pi i)
   =\frac{3N}{\beta_0}e^{10/3}\Lambda^4(\mp\pi i).
\label{eq:(3.17)}
\end{equation}
This is the main result of this paper.

Some remarks are in order. First, in the large-$N$ limit, the contribution of
the W-boson dominates the Borel singularity, i.e.\ the IR renormalon. This can
be seen from the fact that the contribution of the W-boson,
Eq.~\eqref{eq:(3.15)}, is~$O(N)$, while that of the photon
in~Eq.~\eqref{eq:(3.6)} is~$O(N^0)$. This result is in contrast to the argument
in~Ref.~\cite{Anber:2014sda} for small~$N$ that the W-boson does not
contribute to the IR renormalon at all.

Secondly, we note that the following calculation leads us to a completely
different conclusion. If we construct the Borel transform from the
perturbative coefficient for the W-boson of~Eq.~\eqref{eq:(3.12)}, where the
large-$N$ limit is not taken for each perturbative coefficient,\footnote{At
this stage, the large-$N$ limit is taken only for the loop integrands, $L^k$
and~$T^k$, but this limit is not considered after the loop integral, which
gives the additional $N$ dependence. This treatment is not systematic because
only part of the subleading effects is considered. (In fact, the $j\neq0$ terms
in the first line of~Eq.~\eqref{eq:(3.12)} are subleading compared to the $j=0$
term there, as can be seen from the fact that the first term
in~Eq.~\eqref{eq:(3.14)} is exactly the same as the $j=0$ term.) However, we
demonstrate here that even when the integrand is exactly given by the
logarithmic factor as in~Eq.~\eqref{eq:(3.12)}, there is a subtle issue on how
to take the large-$N$ limit.\label{footnote:9}} we obtain
\begin{align}
   &\widetilde{B}(u)_{\text{W-boson}}
\notag\\
   &=-3N\sum_{j=-\infty}^\infty
   \int\frac{d^4p}{(2\pi)^4}\,e^{ip_3 2\pi RNj}
   \left(\frac{e^{5/3}\mu^2}{p^2}\right)^u
   -3\sum_{j=-\infty}^\infty
   \int\frac{d^4p}{(2\pi)^4}\,e^{ip_32\pi Rj}(-1)
   \left(\frac{e^{5/3}\mu^2}{p^2}\right)^u
\notag\\
   &=\frac{3N}{16\pi^2}(e^{5/3}\mu^2)^u
   \left[(q^2)^{2-u}\frac{1}{u-2}
   -2(\pi^2R^2N^2)^{u-2}\frac{{\mit\Gamma}(2-u)}{{\mit\Gamma}(u)}\zeta(4-2u)
   \right]
   -B(u)_{\text{photon}}.
\label{eq:(3.18)}
\end{align}
In this Borel transform, one can see that the pole singularity at~$u=2$
disappears, but a pole singularity at~$u=3/2$ emerges instead. This is a
conclusion completely opposite to the one following from~Eq.~\eqref{eq:(3.15)},
which shows the presence of the singularity at~$u=2$. This peculiar situation
indicates that the large-$N$ limit and the construction of the Borel transform
are not commutable:\footnote{The following remark may be useful.
From~Eq.~\eqref{eq:(1.5)}, the coefficients of the perturbative series $f_k$
corresponding to the Borel transform in~Eq.~\eqref{eq:(3.18)} are given by
\begin{equation}
   f_k=\left.\frac{\partial^k}{\partial u^k}
   \widetilde B(u)_{\text{W-boson}}\right|_{u=0}.
\label{eq:(3.19)}
\end{equation}
Therefore, the contribution of the $j\neq0$ terms in~Eq.~\eqref{eq:(3.18)}
to~$f_k$ is
\begin{equation}
   \frac{3N}{16\pi^2}(-2)(\pi^2R^2N^2)^{-2}
   \left.\frac{\partial^k}{\partial u^k}
   e^{2u\ln(e^{5/6}\mu\pi RN)}g(u)\right|_{u=0},
\label{eq:(3.20)}
\end{equation}
where we defined the $N$-independent function $g(u)$ by
\begin{equation}
   g(u)\equiv\frac{{\mit\Gamma}(2-u)}{{\mit\Gamma}(u)}\zeta(4-2u),
\label{eq:(3.21)}
\end{equation}
and $g(0)=0$, and~$g'(0)=\pi^4/90$. Then, the $j\neq0$ terms give the following
leading-$N$ dependence to the perturbative coefficient $f_k$ with \emph{the
order~$k$ kept fixed}:
\begin{equation}
   \frac{3N}{16\pi^2}(-2)(\pi^2R^2N^2)^{-2}
   k[2\ln(e^{5/6}\mu\pi RN)]^{k-1}\frac{\pi^4}{90}=O(N^{-3}(\ln N)^{k-1}).
\label{eq:(3.22)}
\end{equation}
This is sub-dominant compared to the $j=0$ term in~Eq.~\eqref{eq:(3.18)} that
is~$O(N)$. On the other hand, if one considers the large-order behavior
of~$f_k$ as~$k\to\infty$ \emph{with $N$ kept fixed}, the terms of~$j\neq0$
produce the Borel singularity at~$u=2$, as shown by the second term in the
square parentheses of~Eq.~\eqref{eq:(3.18)}, which is~$O(N)$.}
\begin{equation}
   \sum_{k=0}^\infty
   (\lim_{N\to\infty}f_k)\frac{u^k}{k!}
   \neq 
   \lim_{N\to\infty}
   \left(\sum_{k=0}^\infty f_k\frac{u^k}{k!}\right).
\label{eq:(3.23)}
\end{equation}
We see that this inequality holds especially around $u=3/2$
and~$2$.\footnote{The equality holds in the vicinity of~$u=0$. Thus, if one
defines the Borel transform of the right-hand side by the analytic
continuation of the result around~$u=0$ in the large-$N$ limit, the second
term inside the square brackets of~Eq.~\eqref{eq:(3.18)} vanishes and the
singularity at~$u=2$ follows.} In the present paper, we find the Borel
singularity at~$u=2$ under the procedure where we first determine the
perturbative coefficients in the large-$N$ limit and then construct the Borel
transform. From the perspective of our original subject of how the perturbative
series diverges in the large-$N$ theory, we should adopt this ordering of
operations.\footnote{The first version of the present paper concluded the
singularity at~$u=3/2$ based on the calculation leading
to~Eq.~\eqref{eq:(3.18)}. However, we consider that there are some problems
with this treatment: as noted in~footnote~\ref{footnote:9}, we keep a part of
subleading effects in this calculation, which is not well justified.} We
emphasize that this subtlety is peculiar to the W-boson, which acquires the
twisted momentum.\footnote{If momentum is not twisted, the order counting
in~$1/N$ is straightforward (like in the photon case), and such a subtlety
does not arise. We also note that this subtlety is irrelevant
to~Refs.~\cite{Ishikawa:2019tnw,Ishikawa:2019oga}, where the loop momentum of
the renormalon diagram is not twisted.}

We finally make a comment on an example of the UV-finite quantity which
possesses the renormalon ambiguity corresponding to the Borel singularity
at~$u=2$. The gluon condensate is quartically divergent, as seen
from~Eq.~\eqref{eq:(3.1)}, and it may not be regarded as a physical
observable. However, we may consider (as in~Ref.~\cite{Suzuki:2018vfs}) the
gluon condensate of the gauge field defined by the Yang--Mills gradient
flow~\cite{Luscher:2010iy,Luscher:2011bx}. We can repeat the above analysis for
this perfectly UV-finite quantity, and obtain the same renormalon ambiguity as
the gluon condensate investigated above.

\section{Conclusion}
\label{sec:4}
In this paper we have studied the IR renormalon ambiguity in the gluon
condensate in the $SU(N)$ QCD(adj.) on~$\mathbb{R}^3\times S^1$ with the
$\mathbb{Z}_N$ twisted boundary conditions. In the large-$N$ limit within the
the large-$\beta_0$ approximation, we showed that the Borel transform develops
a pole singularity at~$u=2$. This provides an example that the system in the
compactified space~$\mathbb{R}^3\times S^1$ possesses the renormalon ambiguity
identical to that in the uncompactified space~$\mathbb{R}^4$. This situation
is caused by the W-boson---the gauge field which acquires the twisted KK
momentum due to the twisted boundary conditions---and this is quite different
from the $\mathbb{C}P^{N-1}$ model on~$\mathbb{R}\times S^1$. We hope that the
observation made in this paper can be of relevance to the conjectured
cancellation of the renormalon ambiguity by the instability associated with
the semi-classical bion solution.

\section*{Acknowledgments}
H.S. would like to thank the Theoretical Physics Department of CERN, where part
of this work was carried out, for kind hospitality.
We are grateful to Mohamed M.~Anber, Aleksey Cherman, and Mithat \"Unsal for
valuable comments.
This work was supported by JSPS Grants-in-Aid for Scientific Research numbers
JP18J20935 (O.M.), JP16H03982 (H.S.), and~JP19K14711 (H.T.).

\appendix

\section{$SU(N)$ generators}
\label{sec:A}
We follow the convention in~Chap.~13 of~Ref.~\cite{Georgi:1999wka}. The Cartan
generators in the fundamental representation are taken as
\begin{equation}
   (H_m)_{ij}=\frac{1}{\sqrt{2m(m+1)}}
   \left(\sum_{k=1}^m\delta_{ik}\delta_{jk}-m\delta_{i,m+1}\delta_{j,m+1}\right),
   \qquad m=1,\dotsc,N-1,
\label{eq:(A1)}
\end{equation}
whereas $(N-1)N$ raising and lowering generators are taken as (here $m$, $n$,
\dots\ run from~$1$ to~$N$)
\begin{equation}
   (E_{mn})_{ij}=\frac{1}{\sqrt{2}}\delta_{im}\delta_{jn},\qquad m\neq n,\qquad
   E_{mn}^\dagger=E_{nm}.
\label{eq:(A2)}
\end{equation}
In terms of these generators, the $SU(N)$ algebra reads
\begin{align}
   &[H_m,H_n]=0,
\notag\\
   &[H_\ell,E_{mn}]=(\nu^m-\nu^n)_\ell E_{mn},
\notag\\
   &[E_{mn},E_{pq}]
   =\begin{cases}
   (\nu^m-\nu^n)\cdot H,&\text{when $m=q$ and~$n=p$},\\
   -\frac{1}{\sqrt{2}}E_{pn},&\text{when $m=q$ and~$n\neq p$},\\
   \frac{1}{\sqrt{2}}E_{mq},&\text{when $m\neq q$ and~$n=p$},\\
   0,&\text{otherwise},\\
   \end{cases}
\label{eq:(A3)}
\end{align}
where the $\nu^m$ denote the weights ($(\nu^m)_i=(H_i)_{mm}$; here, no sum is
taken over~$m$) and thus $\nu^m-\nu^n$ are the roots. We note that
\begin{equation}
   \nu^i\cdot\nu^j=-\frac{1}{2N}+\frac{1}{2}\delta_{ij}.
\label{eq:(A4)}
\end{equation}

The above generators are normalized such that
\begin{equation}
   \tr(H_mH_n)=\frac{1}{2}\delta_{mn},\qquad
   \tr(E_{mn}E_{pq})=\frac{1}{2}\delta_{mq}\delta_{np},\qquad
   \tr(H_\ell E_{mn})=0.
\label{eq:(A5)}
\end{equation}

\section{Proofs of Eqs.~\eqref{eq:(3.3)} and~\eqref{eq:(3.9)}}
\label{sec:B}
We start with
\begin{equation}
   \left|\sum_{j\neq0}\sigma_{j,N}\int_0^1dx\,e^{ixp_32\pi Rj}x(1-x)K_\nu(z)\right|
   <\sum_{j\neq0}|\sigma_{j,N}|\int_0^1dx\,x(1-x)K_\nu(z).
\label{eq:(B1)}
\end{equation}
We first note that
\begin{equation}
   |\sigma_{j,N}|\leq\begin{cases}
   1,&\text{for $j=0\bmod N$},\\
   \frac{4}{N},&\text{for $j\neq0\bmod N$},\\
   \end{cases}
\label{eq:(B2)}
\end{equation}
from Eqs.~\eqref{eq:(2.22)} and
\begin{align}
   \left|\dfrac{e^{-i\ell2\pi j/N}-1}{e^{-i2\pi j/N}-1}
   -\ell e^{-i\ell2\pi j/N}\right|
   &=\left|\sum_{n=0}^{\ell-1}\left(e^{-i2\pi j/N}\right)^n
   -\ell e^{-i\ell2\pi j/N}\right|
\notag\\
   &<\sum_{n=0}^{\ell-1}1+\ell=2\ell.
\label{eq:(B3)}
\end{align}
Next, as explained in~Appendix~B of~Ref.~\cite{Ishikawa:2019tnw}, one can show,
for the bounds, that
\begin{equation}
   K_0(z)<\frac{2}{z}e^{-z/2},\qquad K_1(z)<\frac{2}{z}e^{-z/2},\qquad
   \text{for~$z>0$}.
\label{eq:(B4)}
\end{equation}
From these, using $K_2(z)=K_0(z)+\frac{2}{z}K_1(z)$, we have
\begin{equation}
   K_2(z)<\left[\frac{2}{z}+\left(\frac{2}{z}\right)^2\right]e^{-z/2},\qquad
   \text{for~$z>0$}.
\label{eq:(B5)}
\end{equation}

Now, using the above relations, we can proceed as, for instance,
\begin{align}
   &\left|\sum_{j\neq0,j=0\bmod N}
   \sigma_{j,N}\int_0^1dx\,e^{ixp_32\pi Rj}x(1-x)K_0(z)\right|
\notag\\
   &<\frac{2}{(p^2)^{1/2}\pi R}
   \sum_{j\neq0,j=0\bmod N}\frac{1}{|j|}
   \int_0^{1/2}dx\,\sqrt{x(1-x)}
   e^{-\sqrt{x(1-x)p^2}\pi R|j|},
\label{eq:(B6)}
\end{align}
where we have used $z=\sqrt{x(1-x)p^2}2\pi R|j|$ in~Eq.~\eqref{eq:(2.26)}.
Noting that $x/2\leq x(1-x)\leq1/4$ for~$0\leq x\leq1/2$, we have the further
bounds
\begin{align}
   &<\frac{1}{(p^2)^{1/2}\pi R}
   \sum_{j\neq0,j=0\bmod N}\frac{1}{|j|}\int_0^{1/2}dx\,
   e^{-\sqrt{xp^2/2}\pi R|j|}
\notag\\
   &<\frac{4}{(p^2)^{3/2}(\pi RN)^3}
   \sum_{k=1}^\infty\frac{1}{k^3}\int_0^\infty dx\,e^{-\sqrt{x}}
\notag\\
   &=\frac{8}{(p^2)^{3/2}(\pi RN)^3}\zeta(3)=O(1/N^3).
\label{eq:(B7)}
\end{align}

Similarly,
\begin{align}
   &\left|\sum_{j\neq 0\bmod N}
   \sigma_{j,N}\int_0^1dx\,e^{ixp_32\pi Rj}x(1-x)K_0(z)\right|
\notag\\
   &<\frac{8}{N(p^2)^{1/2}\pi R}
   \sum_{j\neq0\bmod N}\frac{1}{|j|}\int_0^{1/2}dx\,\sqrt{x(1-x)}
   e^{-\sqrt{x(1-x)p^2}\pi R|j|}
\notag\\
   &<\frac{4}{N(p^2)^{1/2}\pi R}
   \sum_{j\neq0\bmod N}\frac{1}{|j|}\int_0^{1/2}dx\,
   e^{-\sqrt{xp^2/2}\pi R|j|}
\notag\\
   &<\frac{16}{N(p^2)^{3/2}(\pi R)^3}
   \sum_{j=1}^\infty\frac{1}{j^3}\int_0^\infty dx\,e^{-\sqrt{x}}
\notag\\
   &=\frac{32}{N(p^2)^{3/2}(\pi R)^3}\zeta(3)=O(1/N).
\label{eq:(B8)}
\end{align}
Equations~\eqref{eq:(B7)} and~\eqref{eq:(B8)} imply Eq.~\eqref{eq:(3.3)}
for~$\nu=0$.

In a similar manner, noting Eq.~\eqref{eq:(B5)}, we have
\begin{align}
   &\left|\sum_{j\neq0,j=0\bmod N}
   \sigma_{j,N}\int_0^1dx\,e^{ixp_32\pi Rj}x(1-x)K_2(z)\right|
\notag\\
   &<\frac{8}{(p^2)^{3/2}(\pi RN)^3}\zeta(3)
   +\frac{16}{(p^2)^2(\pi RN)^4}\zeta(4)=O(1/N^3),
\label{eq:(B9)}
\end{align}
and
\begin{align}
   &\left|\sum_{j\neq 0\bmod N}
   \sigma_{j,N}\int_0^1dx\,e^{ixp_32\pi Rj}x(1-x)K_2(z)\right|
\notag\\
   &<\frac{32}{N(p^2)^{3/2}(\pi R)^3}\zeta(3)
   +\frac{64}{N(p^2)^2(\pi R)^4}\zeta(4)=O(1/N).
\label{eq:(B10)}
\end{align}
These imply Eq.~\eqref{eq:(3.3)} for~$\nu=2$. Noting Eq.~\eqref{eq:(B2)},
Eqs.~\eqref{eq:(B7)} and~\eqref{eq:(B9)} imply Eq.~\eqref{eq:(3.9)}.


\end{document}